\documentclass{epl}
\usepackage{epsfig}

\title{Diffusion-limited loop formation of semiflexible polymers:
Kramers theory and the intertwined time scales of chain relaxation
and closing}
\shorttitle{Diffusion-limited loop formation of semiflexible polymers}
\author{S. Jun\inst{1} \and J. Bechhoefer\inst{1}\thanks{E-mail: \email{johnb@sfu.ca}} \and B.-Y. Ha\inst{2}\thanks{E-mail: \email{byha@uwaterloo.ca}}}
\shortauthor{S. Jun \etal}
\institute{
  \inst{1} Physics Department, Simon Fraser University, Burnaby, B.C. V5A 1S6, Canada\\
  \inst{2} Physics Department, University of Waterloo, Waterloo, Ontario N2L 3G1, Canada
\pacs{87.15.He}{Dynamics and conformational changes}
\pacs{82.37.Np}{Single molecule reaction kinetics, dissociation, etc}
\pacs{82.20.Pm}{Rate constants, reaction cross sections, and activation energies}}

\begin{document}

\maketitle

\begin{abstract}
We show that Kramers rate theory
gives a straightforward, accurate estimate of the closing time
$\tau_c$ of a semiflexible polymer that is valid in cases of physical
interest.  The calculation also reveals how the time scales of chain 
relaxation and closing are intertwined, illuminating an apparent 
conflict between two ways of calculating $\tau_c$ in the flexible limit.
\end{abstract}

The looping of polymers is a physical process that allows contact
and chemical reaction between chain segments that would otherwise be too
distant to interact.  Polymer loops are particularly important in
biology:  In gene regulation, looping allows a DNA-bound protein to
interact with a distant target site on the DNA, greatly multiplying
enzyme reaction rates~\cite{Schlief92,Rippe01}.  Similarly, DNA
looping in the 30-nm chromatin fiber may trigger the initiation of
DNA replication at different sites along the DNA by enabling
long-distance interactions~\cite{Jun}.  In protein folding, two
distant residues start to come into contact via
looping~\cite{Thirumalai99,Thirumalai95}.  Measurements of loop formation in
single-stranded DNA segments with complementary ends have also been
used to extract elasticity information (e.g., the sequence-dependent
stiffness of single-stranded DNA~\cite{Libchaber00}).

Despite its
importance and despite considerable theoretical effort, there are
relatively few analytical results concerning the dynamics of loop
formation.  Even for the simplest case of an ideally flexible
polymer with no hydrodynamic effects (or simply a Rouse chain), there are two rival theoretical approaches that lead to contradictory results:  Szabo, Schulten, and Schulten (SSS) conclude that the time for a loop to form (``closing time" $\tau_c$) should scale for moderately large polymer lengths $L$ as $\tau_c \sim
L^{3/2}$~\cite{SSS80}, while Doi,  applying Wilemski-Fixmann (WF)
theory~\cite{WF74}, finds $\tau_{Doi} \sim L^2$~\cite{Doi75}.  The
discrepancy between the two continues to spur debate~\cite{Pastor96,Portman03}.  For the important case of stiff
chains~\cite{Schiessel01, Schiessel02}, where the polymer length $L$ is not too much longer than the persistence length $\ell_p$, only limited numerical results are known [see, for example,~\cite{Alexei00, Dua} and references therein].
The main difficulty arises from the interplay between two seemingly 
distinct processes: chain relaxation and chain closure.  This  
interplay is unique to a polymeric system and originates from the 
chain connectivity of a polymer immersed in a noisy 
environment.

In this article, we argue that
Kramers' rate theory~\cite{Kramers40,Hanngi90} applies to the most
physically relevant cases and leads to analytical results for
$\tau_c$.  We capture, for the first time, that there is a minimum loop-formation time for chain lengths of approximately  3-4  $\ell_p$.  Roughly speaking, shorter chains require too much energy relative to the thermal energy $k_B T$, while longer 
chains need to search too many conformations for ends to ``find" each other.  We
also show that consideration of the requirements for Kramers theory
to apply leads one naturally to identify different regimes governing
the closing time $\tau_c$.  This classification shows how the physics
of chain relaxation is intertwined with that of chain closing and clarifies the above-mentioned controversy between the SSS and Doi approaches to loop-formation dynamics.

Consider a chain of length $\ell \equiv L/\ell_p$ with two ends that react
when first brought within a distance $a$ of each other
(``diffusion-limited" loop formation dynamics).  We apply Kramers
rate theory, viewing the process as a noise-assisted ``tunneling"
over a potential barrier.  After first presenting the straightforward
calculation, we then consider carefully its domain of applicability, 
followed by a scaling description of loop formation outside this 
domain.

\begin{figure}[t]
\centering
\epsfig{file=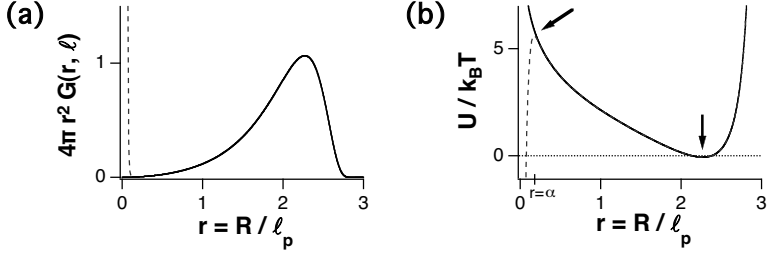, width=4.0in}
\caption{(a) The radial distribution density $P(r,\ell=3)$.  The dashed
line shows the effect of a short-range interaction between the two
polymer ends.  (b) The resulting effective potential of the chain.  Arrows denote the top and bottom of the effective potential well, as used in the Kramers calculation.}
\label{fig:end2end}
\end{figure}

The basic idea is to project the internal degrees of freedom of the polymer chain onto a single ``reaction coordinate" $r \equiv R/\ell_p$, with $R$ the end-to-end distance of the chain.  The reduced, one-dimensional dynamics then obey a Langevin equation of the form
\begin{equation}
\label{eq:langevin}
\frac{dr}{dt}=-\frac{D}{k_B T}\partial_r U(r,\ell) + \xi(t) ,
\end{equation}
where $D = 2D_0$ is twice the
diffusion coefficient of a monomer (both ends diffuse
\cite{Pastor96}) and $\xi(t)$ represents Gaussian white noise:
$\langle \xi(t) \rangle = 0$ and
$\langle\xi(t)\xi(t')\rangle = 2D \delta(t-t')$, with $\langle ... 
\rangle$ a thermal average.  The dynamics are governed by an 
effective potential $U(r,\ell)$.  Strictly speaking, this description is valid for a chain in local equilibrium, for which we can write 
\begin{equation}
\label{eq:potential}
U(r,\ell) = -k_B T \ln P(r,\ell)  ,
\end{equation}
where $P(r,\ell) = 4\pi r^2 G(r,\ell)$ is the radial distribution function
of end-to-end distances $r$ of a polymer of length $\ell$ and $G(r,\ell) 
\equiv G(|\mathbf{r}_{\ell}-\mathbf{r}_0|; \ell)$, the angle-averaged 
distribution function for the end-to-end vector 
$\mathbf{r}=\mathbf{r}_{\ell}-\mathbf{r}_0$.  We assume isotropic chemical 
interactions
between end monomers, so that end binding can be modeled by adding to
$U$ a smooth short-range potential $f(r/\alpha)$, with $\alpha \equiv
a/\ell_p$ the interaction range.  A typical distribution function and
resulting effective potentials are shown in fig.~\ref{fig:end2end}.
 
Because polymers -- whatever their stiffness -- have a most probable
end-to-end separation (radius of gyration), there is a local minimum
in the effective potential at $r_b$ (bottom), which is $\sim \ell$ in
the stiff-chain limit and $\sim \sqrt{\ell}$ in the flexible-chain
limit, neglecting self-avoidance effects.   Also notice the barrier 
to chain closing at $r_t \approx \alpha$ (top), which is created by 
the balance of chain entropy and bending energy as implied by 
$U(r,\ell)$. The short-range attractive potential then rounds off the 
barrier.  The resulting effective potential has thus the qualitative 
form often assumed in Kramers-rate calculations.

In the limit of strong damping \cite{intermKr}, the time needed to
tunnel over the barrier (mean first-passage time), calculated using Kramers rate theory, is
\begin{equation}
    \tau_{Kr} = \left[
    \frac{D}{k_B T} \frac{\omega_{t} \omega_{b}}{2\pi} \exp\left(-
    \frac{\Delta U}{k_B T}\right) \right]^{-1} ,
\label{eq:tau_c}
\end{equation}
where the well curvatures {\bf $\omega(r) = \frac{1}{\ell_p}
\sqrt{\partial_{rr} U(r,\ell)}$} are evaluated at the top and
bottom of the effective potential $U(r,l)$.  The exponential factor is
\begin{equation}
     \exp \left(-\frac{\Delta U}{k_BT} \right) = 
\frac{P(r_t,\ell)}{P(r_b,\ell)} \\
      \simeq \frac{\alpha^2 G(0,\ell)}{r_b^2
G(r_b,\ell)}, \quad (\alpha \ll 1) .
\end{equation}
We find the surprisingly simple result \cite{barrier}
\begin{equation}
   \tau_{Kr} (\ell) =\mathcal{C} \frac{1}{\alpha D}  \frac{\ell_p^2}{G_0(\ell)},
\label{eq:firstpass3}
\end{equation}
with $G_0 \equiv G(0,\ell)$ and with $\mathcal{C} \left( r_b, G(r_b,\ell) \right) = \sqrt{2} \pi r_b^2
G(r_b,\ell) \big/\left( {6 \over r_b^2}-{G''(r_b,\ell) \over G(r_b,\ell)}
\right)^{1/2}$ a dimensionless prefactor that is 
practically a constant for all $\ell$ (see below).

Eq.~\ref{eq:firstpass3} shows that the closing time may be
estimated using the static distribution $G(r,\ell)$.  Unfortunately, no
analytic expression for $G(r,\ell)$ has been found that is accurate for
all $r$ and $\ell$, and one must make 
a pastiche of approximations. For $r = 0$ and $\ell < 10$, we use an approximation  for a wormlike chain derived by Shimada and Yamakawa \cite{SY84,Yamakawa97}:
$G_0(\ell) = (896.32/\ell^5) \exp(-14.054/\ell + 0.246 \, \ell)$.  Note here that the $1/\ell$-term in the exponent solely arises from bending energy, while the rest comes from chain fluctuations around the lowest-energy conformation.   For $r = 0$ and large $\ell$,
we use an interpolative formula due to Ringrose \etal~that blends SY
with the result for a freely jointed chain, $G_0(\ell) \sim \ell^{-3/2}$
\cite{Ringrose99}.  Near $r = r_b$, we use an approximation derived
by Thirumalai and Ha (TH)\cite{ThirHa98}, valid to 10\%:
$G(r,\ell) = n(\ell) [1-({r \over \ell})^2]^{-{9 \over 2}} \exp \big\{-{3\ell \over 4} {1\over [1-(r/\ell)^2]}\big\}$, with
the normalization factor $n(\ell)$ fixed by requiring $\int_0^\ell 4\pi r^2
G(r,\ell) dr = 1$.  Note that a more accurate but more complicated
expression recently derived by Winkler \cite{Winkler03} gives
essentially the same results.   Using TH, we find that the
dimensionless prefactor $\mathcal{C}(\ell)$ of eq.~\ref{eq:firstpass3}
is $\mathcal{O}(10^{-1})$, varying less than a factor of 2 over $0 <
\ell < \infty$.

\begin{figure}[t]
\centering
\epsfig{file=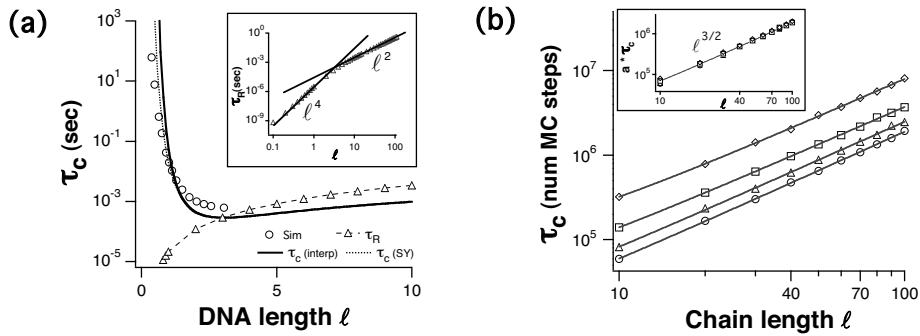, width=4.8in}
\caption{Closing time $\tau_{c}$ vs.~chain length.  (a) BD simulation \cite{Alexei00} (empty circles) and Kramers theory (eq.~\ref{eq:firstpass3}) are shown.  For direct comparison, we used the same parameters as in the Ref.~\cite{Alexei00} (bead size = 3.18 nm for $D=2 D_0=1.54 \times 10^{-11}$m$^2$/s and $\alpha=0.1$) with $\ell_p$ = 50nm.  For $G_0(\ell)$, we used the SY result~\cite{SY84} and an interpolation~\cite{Ringrose99} (see text).  Relaxation times $\tau_R$ for these parameters are also shown (triangular symbols), with the $\ell^4$ and $\ell^2$ scaling regimes apparent in the inset.
(b) Single-``particle" MC simulations of $\tau_c$ with the
potential ${U/k_BT} = -\log[P(r,\ell)]$ taken from Fig.~\ref{fig:end2end}b.
Here, $\tau_c$ is a first-contact time averaged over about 2000
realizations of the initial position randomly selected from $P(r,\ell)$.
We have chosen $\alpha=0.25, 0.5, 0.75, 1.0$.   As expected,  $\tau_c
\sim \frac{\ell^{3/2}}{a}$ (inset).}
\label{fig:THvsSIM}
\end{figure}

In fig.~\ref{fig:THvsSIM}a, we plot the $\tau_{Kr}(\ell)$
that results from eq.~\ref{eq:firstpass3}, using the various
approximations to $G(r,\ell)$ discussed above.  The solid curve uses the
Ringrose expression for all $\ell$, while the dashed curved uses SY for
small $\ell$.  The two curves compare well with recent simulations using
parameters appropriate to double-stranded DNA \cite{Alexei00}.  Note
that the material parameters of the simulation were used (see
caption).  Considering the
heuristic nature of the arguments, the agreement is excellent.

One striking feature of the plot of $\tau_{Kr}(\ell)$ is the existence of a
minimum at $\ell \approx 3.4$, where
\begin{equation}
    \tau_{Kr}^* = 0.78
\frac{\ell_p^3}{D_0 a} .
    \label{eq:tau_min}
\end{equation}
In eq.~\ref{eq:tau_min}, the prefactor $0.78$ is calculated by Monte
Carlo simulation of $G(r,\ell)$, in units of seconds,  and is about
10\% less than the prefactor obtained using the TH approximation.  As
mentioned above, the existence of a minimum in $\tau_{Kr}$ reflects a
balance between the energy of bending and the entropy of
conformations that must be searched for two ends to meet.

For the above Kramers-rate calculation to hold so that $\tau_{Kr}$
equals $\tau_c$, three conditions must be satisfied:  The damping
must be sufficiently strong; the barrier height $\Delta U$ must be
large compared to $k_B T$; and the global chain relaxation time $\tau_R$,
a characteristic time scale for chain deformation, must be much shorter than the Kramers time $\tau_{Kr}$.

\begin{figure}[t]
\centering
\epsfig{file=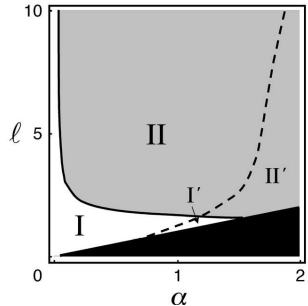, width=1.6in}
\caption{Scaling regimes in the $\ell$-$\alpha$ for DNA (see text).
Region I is the Kramers regime, with $\tau_c > \tau_R$; Region II is the
dynamic-fluctuation regime.  In the primed regions to the right of
the dashed line, $\Delta U/k_BT < 1$.  The black region is
unphysical: $a > L$.}
\label{fig:map}
\end{figure}

The first condition is normally satisfied for molecules
in solution \cite{intermKr}.  For the second, since there is a minimum
in the effective potential at $r_b$, we require that $\alpha$ be
sufficiently less than $r_b$ so that the barrier height is large.
The condition $\Delta U / k_BT = 1$ is shown in fig.~\ref{fig:map} as
a dotted line in the $\ell-\alpha$ parameter plane, using a diffusion
constant appropriate to double-stranded DNA.  To the left of the
dashed line, the barrier height is larger than $k_B T$.

The third condition, $\tau_R \ll \tau_{Kr}$, is more subtle and requires
discussion.  In using a ``one-particle" description of chain closing
dynamics, we are assuming that all internal degrees of freedom of the
polymer chain have relaxed.  As a result, the end-to-end distance is 
the only dynamic variable (Cf. eq.~\ref{eq:langevin}).  This 
assumption of local equilibrium allows one to apply the equilibrium 
distribution function $G(r,\ell)$
and implies that the effective potential derived from $G$ is time
independent.  If the chain relaxation times are too long, the
potential effectively becomes time dependent and has to be obtained
self-consistently, along with the motion of the internal modes.  We
thus compare the scaling behavior of $\tau_R(\ell)$ with $\tau_{Kr}(\ell)$
and $\tau_c(\ell)$ in both the flexible ($\ell \gg 1$) and stiff-chain ($\ell 
< 1$) limits.

In the flexible limit, we can use the Rouse model to estimate the
longest relaxation time, which gives $\tau_R \sim \ell^2$, in units of
the basic time scale $\ell_p^2/D$. By contrast, at large $\ell$ eq.~\ref{eq:firstpass3} gives $\tau_{Kr} \sim \ell^{3/2} / \alpha $.  (This
is just the result of SSS \cite{SSS80,Pastor96} and has been confirmed by single-``particle" simulations---see fig.~\ref{fig:THvsSIM}(b) and the caption.)  
Thus, when $\ell > 1/ \alpha^2$, the third condition is violated and the Kramers
calculation does not hold.  Nonetheless, we can still estimate the
upper-limit of $\tau_c$:  The closing time is at most a time 
necessary for the slowest ``random walker" to travel, with 
diffusion constant that of the entire chain $D_{chain}\sim D/l$, 
the end-to-end distance $r$.  Since $r \sim \sqrt{\ell}$, we have $\tau_c < r^2/D_{chain} \sim \ell^2/D \sim \tau_R$.  In other words, when the 
third condition does not hold, $\tau_c$ is not $\tau_{Kr}$ but is set 
by the Rouse time $\tau_R$.

In the stiff limit, the physics is
dominated by the bending energy ${E_b}$ of a rod 
\cite{Wilhelm96}, leading simultaneously to faster relaxation times 
$\tau_R$ and higher energy barriers, which implies that the Kramers 
calculation should be valid.
To see this, recall that the bending energy of an elastic rod is, by
symmetry, proportional to the square of the rod curvature.  Since the
lowest energy corresponds to a uniform curvature of radius
$\mathcal{R}$, the bending energy near the rod limit $E_{b}/k_BT = 
\frac{1}{2} \ell_p L/\mathcal{R}^2 \sim 2 \ell_p (L-R)/L^2$.

Thus, if we track the relative separation $R$ of the
endpoints of a thermally excited rod, it behaves like a particle
subject to a constant restoring force $f_c = 2k_BT \ell_p/L^2$.  The
appropriate Langevin 
equation for ${\bf R}$ is then of the form
$\dot{\bf R} + \frac{D_{chain}}{k_BT} f_c \frac{{\bf R}}{R}= 
\vec{\xi}_{chain}(t)$, with $\vec{ \xi}_{chain}(t)$  the random force.  This 
implies that the time to relax a distance of order $L$ is $k_BT L 
/D_{chain} f_c  = L^3/2\ell_pD_{chain}$.    Since the rod moves 
coherently, the diffusion coefficient of the chain
$D_{chain} \sim D b/L$, leading to $\tau_R \sim \frac{L^4}{2\ell_p b 
D}$, where $b$ is the monomer size.  As a result, for $L < l_p$, the 
third condition ($\tau_R \ll \tau_{Kr}$) is always satisfied: the
lower-limit of $\tau_c$ is given by a time scale for a random walk to
travel a distance $R$$\sim$$L$, thus $\tau_{Kr}$$\sim$$\tau_c >
R^2/D_{chain}$$\sim$$L^3/b D > L^4/\ell_p b D$$>$$\tau_R$.

To summarize, $\tau_R \sim L^4$ for $\ell < 1$ and $\sim L^2$ for $\ell
\gg 1$.  Thus, for large enough $\ell$,  $\tau_R$ becomes larger than
the Kramers estimate \cite{Harnau97}, as shown in
fig.~\ref{fig:THvsSIM}a and in the inset.  In fig.~\ref{fig:map}, we
plot $\tau_R(\ell) = \tau_{Kr}(\ell)$ in the $\ell-\alpha$ plane.  The white
area is Region I (Kramers Regime), where $\tau_{Kr} > \tau_R$, and
therefore $\tau_c \sim \tau_{Kr}$.  The shaded area is Region II
(``dynamical fluctuation" Regime, see below), where
$\tau_{Kr} < \tau_R$, and therefore $\tau_c \sim \tau_R$.  Areas I'
and II' show where $\Delta U < k_B T$.  The black region, defined by
$\alpha > \ell$, is unphysical.

In Region II, the relaxation and closing processes are coupled.  In this case,
one may have to solve an $N$-particle diffusion problem, subject to a
boundary condition that is difficult to impose~\cite{WF74, Doi75,
Pastor96}.  Nevertheless, much insight can still be obtained from the
simple scaling analysis of random walks given above.  In this view, a
chain can close because the two ends randomly meet each other
while freely relaxing.  The existence of such a regime, where $\tau_c
\sim \tau_R$, is a unique feature of flexible chains
(fig.~\ref{fig:map}) that we denote the ``dynamical fluctuation"
regime---the dynamic fluctuation $\delta R(t) \equiv \sqrt{\left< [ R(t)-R(0) ]^2 \right>}$ grows up to $R$ as $t \rightarrow \tau_R$ and thus can assist 
chain closing.  For a Rouse chain, $\delta R(t)$ can be given as a sum of Rouse modes~\cite{deGennes79} and, in our simple scaling analysis, $\tau_c$ can be inferred by analyzing this.  Short-time behavior of $\delta R(t)$ reflects the internal motion and varies as $\delta R(t) \sim \sqrt{t}$ for $t \ll \tau_R$.  We, however, argue that this will not appreciably influence $\tau_c$, as $\delta R(t) \rightarrow R$ only when $t \rightarrow \tau_R$.  In other words, $\tau_c$ is governed by the slowest mode and our assertion of $\tau_c \sim \tau_R$ will not be invalidated by the internal motion, which is important at time scales much smaller than $\tau_c$ (or $\tau_R$).
 In the stiff-chain limit, this 
dynamical fluctuation regime
disappears.  Note that the boundaries between Regions I and II are
not sharp but are crossovers.  Loop-formation kinetics in the
crossover area will likely combine aspects of both regimes, as
indicated in recent simulations \cite{Pastor96} and by results that
show that
$\tau_{SSS}$ and $\tau_{Doi}$ are respectively lower and upper bounds
for $\tau_c$ \cite{Portman03}.  Similarly, based on their BD
simulation results, Podtelezhnikov \etal~\cite{Alexei97}
suggested that $\tau_c \simeq \tau_R/\alpha$ near the boundaries.

Our discussion has neglected hydrodynamic effects and excluded-volume
interactions.  Both can influence chain relaxation and closing
simultaneously.  The hydrodynamic effect will not change $\tau_{Kr}$, since it is a function of the equilibrium distribution $G(r, l)$.  However, the hydrodynamic interaction tends to promote chain
relaxation (e.g. in the Zimm model, $\tau_R \sim \ell^{3/2}$, in contrast to $\tau_R \sim \ell^2$ in the Rouse model considered here~\cite{deGennes79}) by increasing the mobility of the chain, resulting in a wider Kramers regime than implied by Fig.~\ref{fig:map}.  On the other hand, the excluded-volume interaction both decreases $D_{chain}$ and reduces $G_0$ \cite{deGennes79,Chen02}.  But  for loops of just a few persistence lengths, which are the most physically relevant (see below), both effects are expected to be minor. A final caveat is that we have assumed isotropic binding interactions.  While mathematically simpler and relevant to simulations \cite{Alexei00}, most real polymers have
directional bonding.  In the Kramers calculation, this would modify
$G_0(\ell)$.

The Kramers calculation holds in Region I of the
$\ell-\alpha$ parameter space shown in fig.~\ref{fig:map}.  What are the
physically relevant values of $\alpha$ and $\ell$?  The interaction
distance $a = \alpha \ell_p$ will be the thickness of the polymer, or
less.  For polymers of biological interest, the persistence length
will be typically at least this size and often much larger.  For
example, for double-stranded DNA, the monomer size is 0.34nm while the
persistence length is 50 nm.  For chromatin, the thickness
is 30 nm, comparable to its persistence length~\cite{Dekker}. Thus, we generally expect $\alpha < 1$ and sometimes $\alpha \ll 1$.

What are
the relevant values of $\ell$?  Although polymers in principle may have
any length, the existence of a minimum closing time $\tau_{Kr}^*$
(eq.~\ref{eq:tau_min}) leads one to speculate that where looping is
biologically relevant, polymer lengths near $\ell \approx 3-4$ might be
favored because they minimize $\tau_c$.  In this regime, the Kramers
calculation will be valid, for small $\alpha$.  Thus, biological
selectivity may arise from a physical mechanism.  For example, a
recent study of Jun \etal~\cite{Jun} on DNA replication noted that
the typical spacings between replication origins in early embryo
\textit{Xenopus} are 3-4 times the $\ell_p$ of chromatin,
the DNA-protein complex present during replication.  It is then
natural to speculate that origins are related by looping and that the
spacing may be selected to maximize the contact rate of origins,
optimizing replication efficiency.

In conclusion, we have shown
that Kramers rate theory gives a straightforward estimate of the
closing time of a semiflexible polymer.  Although phenomenological,
the calculation explains the existence of a minimum closing time and
accurately reproduces numerical simulations.  Moreover, considering
the requirements for the calculation to hold shows how the
intertwining of the relaxation time with the closing time explains
the apparently conflicting results for $\tau_c$ (SSS and Doi).
Fortunately, the physically relevant cases are precisely the ones
where the Kramers calculation is expected to hold and may even be
selected biologically through evolution.

\acknowledgments
This work was supported by the Natural Science and Engineering Research Council of Canada
(NSERC).  One of us (Jun) acknowledges the hospitality and financial
support of Pu Chen during his visit to Waterloo.  We are grateful
to J. Chen, B. Cherayil, A. Dua, and M. Wortis for helpful discussions and to H. Imamura for help with MC simulations.  We also thank A. A. Podtelezhnikov for kindly sending us the BD simulation data.


\begin{thebibliography}{0}
\bibitem{intermKr}
For intermediate-to-strong damping, the Kramers time $\tau_{Kr}$ is given by
$\tau_{Kr}^{-1}= \frac{D}{k_BT} \frac{\omega_b \, \omega_t}{\pi} \,  e^{-\frac{\Delta U}{k_BT}} \Big/ { \left( 1 +  \sqrt{1 + \frac{4mD^2 \, \omega_t^2}{(k_BT)^2}}\right)}.$
The correction term $\frac{4mD^2 \, \omega _t^2}{(k_BT)^2}$ is
$\approx 10^{-7}$ for DNA monomers and can be neglected,
justifying our use of the strong-damping limit
(eq.~\ref{eq:tau_c}).
\bibitem{barrier}  For a smooth short-range potential of
range $\alpha$, the curvature at the top must be $\sim 1/\alpha$ by
dimensional analysis.  We note that many simulations assume reaction 
upon first passage through the distance $\alpha$.  Despite the 
seeming difference between our Kramers' approach and simulations that 
track the time for particle ends to first pass through the $r=\alpha$ 
sphere, 
the ``particle" (in a single-particle picture) in both cases 
is not allowed to equilibrate within the reactive region $r \approx 
\alpha$.  Thus, in each case, one expects $\tau_c \sim 1/\alpha$ for 
$\alpha \ll 1$ (Cf. eq.~\ref{eq:firstpass3}).  If we had assumed 
kinetic-limited looping, then the particle would sample most of the 
reactive region, resulting in $\tau_c \sim 1/\alpha^2$~\cite{SSS80}.
\end{thebibliography}
\end{document}